# Decoupling Graphene from SiC(0001) via Oxidation

S. Oida, F.R. McFeely, J.B. Hannon, R.M. Tromp, M. Copel, Z. Chen, Y. Sun, D.B. Farmer and J. Yurkas
[1]*IBM Research Division, T.J. Watson Research Center, Yorktown Heights, NY 10598*

When epitaxial graphene layers are formed on SiC(0001), the first carbon layer (known as the "buffer layer"), while relatively easy to synthesize, does not have the desirable electrical properties of graphene. The conductivity is poor due to a disruption of the graphene π-bands by covalent bonding to the SiC substrate. Here we show that it is possible to restore the graphene π-bands by inserting a thin oxide layer between the buffer layer and SiC substrate using a low temperature, CMOS-compatible process that does not damage the graphene layer.

Following its experimental realization by Novoselov *et al.*[1], graphene, one, or a very few, layers of carbon in hexagonal $sp^2$-hybridized sheets, has been the subject of intensive investigation. Its unique electronic properties[2,3] have attracted great interest owing to potential applications in nanoelectronics[4,5]. Graphene films can be produced in a variety of ways, e.g., exfoliation of samples from pyrolytic graphite[1], chemical vapor deposition (CVD)[6–8], or sublimation of Si from SiC(0001) substrates[9–12]. From the standpoint of compatibility with current device fabrication processes, Si sublimation from SiC(0001) is particularly appealing: Wafer-sized graphene films of controlled thickness can be grown directly on a semi-insulating substrates. However, graphene growth on SiC(0001)[13], either by sublimation of Si or by carbon CVD, has a serious drawback. The first graphene layer, while easy to grow uniformly, is non-conductive[14,15]. Thus from an electronic point of view, this layer is not graphene at all, but rather a "buffer layer" on which additional, electrically-active graphene must be grown. Band structure measurements by angle-resolved photoemission spectroscopy[16] and first-principles calculations[14,15] show disruption of the buffer layer π-bands by strong covalent bonding to the SiC substrate.

Recently, it was shown that annealing in hydrogen at temperatures above 600 °C can decouple the buffer layer from the the SiC substrate, resulting in the appearance of the graphene band structure[17]. Here we describe a low-temperature oxidation process that accomplishes the same decoupling. When the buffer layer is exposed to oxygen at 250 °C, an oxide layer of 3 Å is formed between the buffer layer and the SiC(0001) substrate. Surprisingly, this ultra-thin layer is sufficient to decouple the buffer layer from the substrate, restoring the π-band structure characteristic of free-standing graphene. We correlate the existence of graphene-like bands with the appearance of the plasmon in electron energy loss spectroscopy (EELS).

Although it perhaps seems counter intuitive to attempt to improve the conductivity of a structure by oxidation, the formation of a $SiO_2$ decoupling layer between the graphene and the SiC substrate has a fair amount of *a priori* thermodynamic and kinetic plausibility. The free energy of formation of $SiO_2$ is more negative than that of $CO_2$ by approximately 100 kcal/mole at 500 K. Thus if a buffer layer / SiC structure were oxidized under such conditions so as to achieve thermodynamic equilibrium, essentially all of the oxygen reacted would be in the form of $SiO_2$. Of course, this offers no guarantee that the desired structure can be synthesized, since equilibrium is not achievable under practical oxidation conditions. In fact, the equilibrium products of SiC oxidation are undesirable, as graphitic carbon, presumably highly disordered, would be produced in equimolar amounts to the $SiO_2$. Instead the ideal to be sought is a kinetic regime in which graphene remains inert to the oxidant (e.g. $O_2$), SiC is oxidized to produce sufficient $SiO_2$, and the carbon liberated from the oxidation of SiC is oxidized and carried away. While these requirements appear quite stringent, graphene is well known for its chemical inertness, and, as we show, no more than about a monolayer of $SiO_2$ is required to decouple the film from the substrate. In addition there is precedent for oxidation selectivity such as we desire between the graphene and the nascent carbon formed by SiC oxidation: when carbon nanotubes are grown from alcohol precursors it is believed that one of the roles of the oxygen is to scavenge any amorphous carbon formed in the pyrolyitc process[18]. In what follows we demonstrate all three of these requirements can be met by a process of low temperature, high pressure oxidation. By this means, the buffer layer can be electronically decoupled from the SiC substrate, restoring the π-bands to a substantially unperturbed condition.

The synthesis of epitaxial graphene layers on SiC(0001) was carried out in an ultra-high vacuum system equipped with low-energy electron microscopy (LEEM)[11–13]. The graphene layers were formed at elevated temperature while the surface was imaged with LEEM. Details on the preparation of clean, flat SiC(0001) samples are given elsewhere[11,12]. Two different synthesis approaches were used, yielding essentially identical results. In the first process, the sample is annealed above 1300 °C in a background pressure of disilane until the SiC decomposes, creating 1-3 ML of carbon in a controlled manner. In the second process, a small amount of ethylene (e.g. $1 \times 10^{-7}$ Torr) is added to the disilane background below the temperature at which SiC decomposes. A buffer layer film limited to a single carbon layer can be formed with this CVD approach. This latter method has the advantage that the CVD process is self limiting, yielding a single graphene buffer layer, with no possibility of producing additional graphene, which would complicate the analysis of the experiments. However, the nucleation rate of the graphene buffer layer during CVD growth is difficult to control, and the domain size of the CVD films can be significantly smaller than that of films grown by thermal decomposition. After synthesis, the graphene layer thickness was verified using the LEEM reflectivity method developed by Hibino *et al.*[19]. EELS was used to monitor the integrity of the π-bands. The LEEM instrument employed for these experiments includes an energy filter, enabling us to obtain EELS spectra *in situ* from the same area of the surface that is imaged[20]. A focused ion beam (FIB) system was used to mill out alignment marks on the SiC substrate before graphene synthesis, which allowed us to obtain EELS spectra and images from a specific area of the surface, remove the sample from the LEEM chamber for oxidation, return it to the LEEM chamber and collect LEEM images and EELS spectra from exactly the same area of the sample. For reference, an EELS spectrum from a thick exfoliated graphene flake placed on SiC(0001) is shown in Fig. 1.

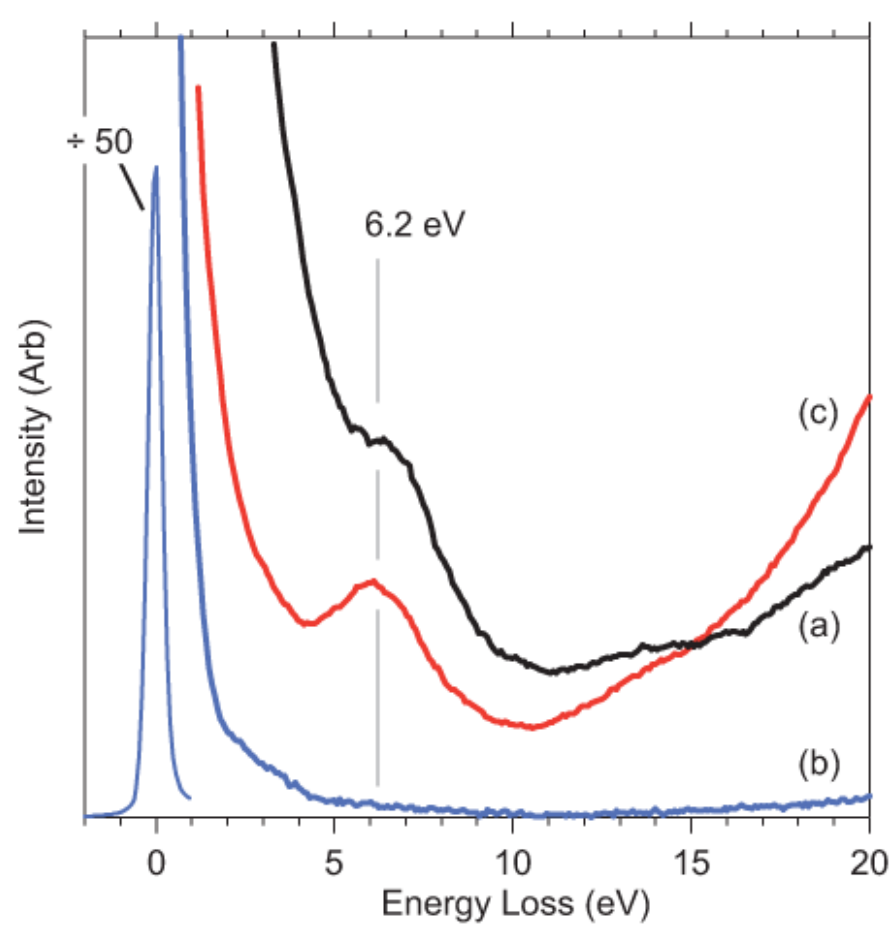

FIG. 1: Electron energy loss spectra recorded from (a) a thick graphite flake placed on SiC(0001) (black), (b) a graphene buffer layer on SiC(0001) before oxidation (blue), and (c) the same buffer layer after oxidation (red). All spectra were recorded using 33 eV electrons at near-normal incidence ($q \sim 0$).

The incident electron energy was 33 eV and the scattering geometry was such that both the incident and scattered beams were approximately normal to the surface ($q$~0). The feature near 6.2 eV loss energy corresponds to the surface π-plasmon of graphite[21–23]. A spectrum obtained under identical scattering conditions from a graphene buffer layer synthesized on SiC via CVD is also shown. As expected, owing to the disruption of the graphene π-bands, no plasmon loss features are observed, confirming that the electronic structure of the covalently-bonded buffer layer is different from that of graphene.

Following LEEM image collection and selected area EELS characterization, the sample was removed from the LEEM and atomic force microscopy (AFM) images were obtained from the same area in which the EELS spectra were obtained. It was then introduced into an oxidation chamber connected to an x-ray photoemission (XPS) spectrometer. The sample was oxidized in 1 atm of $O_2$ at 250 °C for 5 s. The effect of the oxidation on the sample was characterized using XPS, as illustrated in Fig. 2.

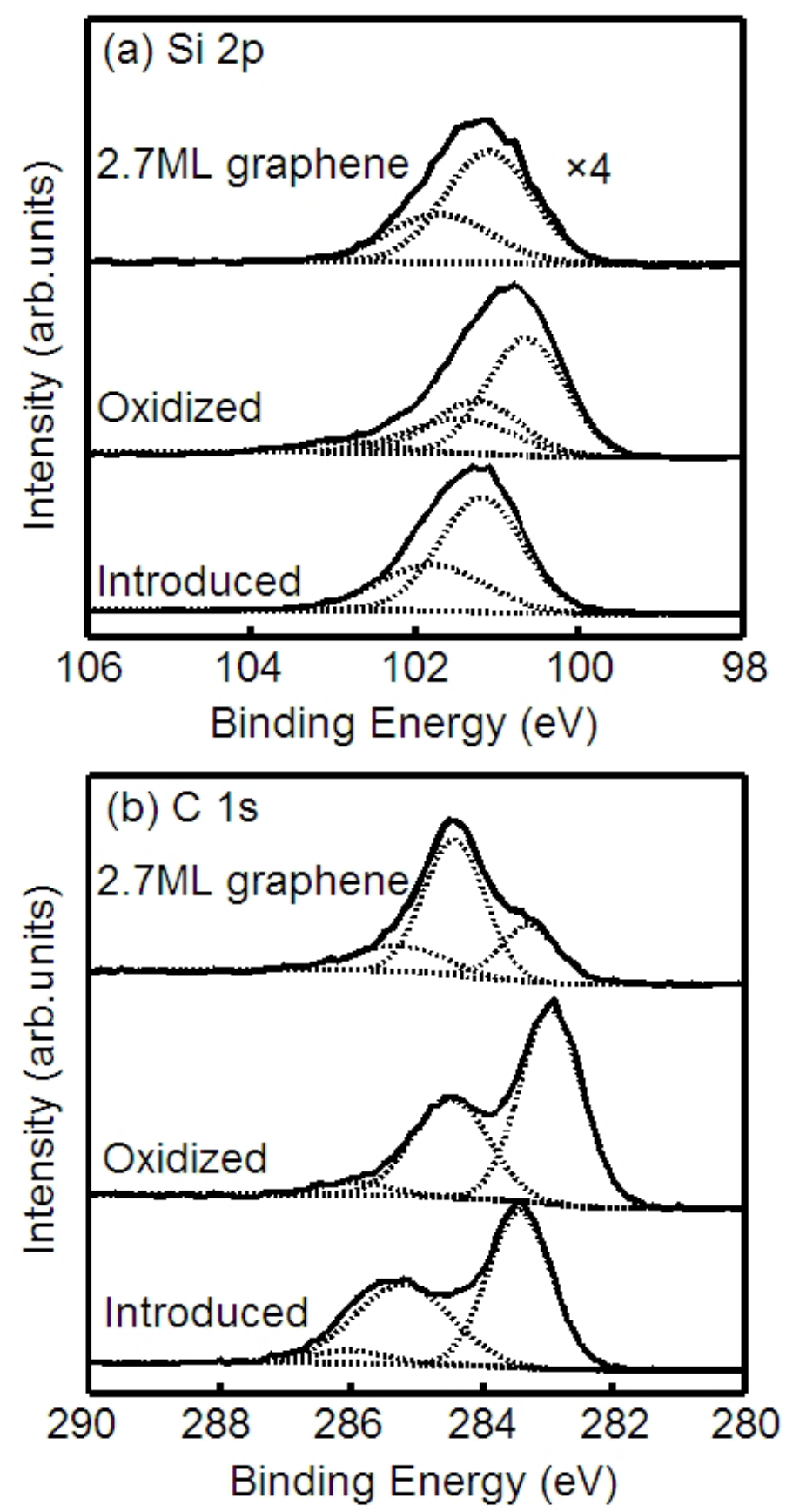

FIG. 2: XPS spectra of the (a) Si 2*p* and (b) C 1*s* core levels from a buffer layer grown on SiC(0001). The bottom spectrum in each panel is from the buffer layer before oxidation. The middle spectra are from the buffer layer after oxidation. The top spectra are from a thick graphene film grown on SiC(0001) via high-temperature sublimation.

In each panel, the bottom spectra show the Si 2*p* and C 1*s* core levels of the buffer layer sample after growth but prior to oxidation. The Si 2*p* region shows a single peak corresponding to Si in SiC. (Some slight tailing to higher binding energy is observed due to a small amount of oxide present on this airexposed sample.) The C 1*s* region shows a peak at 283.42 eV binding energy corresponding to carbidic carbon and a second peak shifted to higher binding energy by 1.81 eV, corresponding to buffer layer carbon. After oxidation, the spectra in the middle rows are obtained. The Si 2*p* spectrum shows a weak satellite feature at higher binding energy, which we attribute to the formation of oxidized Si. (The O 1*s* spectrum, not shown, confirms the uptake of oxygen by the system.) From the area of the weak satellite feature, we estimate the thickness of this oxidized layer, analyzed as $SiO_2$, to be surprisingly thin: no more than about 3 Å. Quantitative ion scattering measurements to determine the oxide thickness more precisely are described below. For these oxidation conditions, the growth of the oxide saturates within seconds and there is little qualitative difference between samples oxidized for 5 s or 1 hr. In the C 1*s* spectrum, the graphene buffer layer peak has shifted by approximately 0.26 eV towards lower binding energy with respect to the carbidic carbon peak. This is hardly surprising, since, as we shall show below, the valence electronic structure of the graphene layer has undergone a dramatic change. However we note that the graphenic carbon intensity is unchanged from the unoxidized sample, indicating that within sensitivity of the XPS measurements (about 5%) the graphene layer is not chemically attacked by the oxygen, and any carbon liberated from the SiC via oxidation is removed from the sample by the oxygen, either as CO or as $CO_2$. The latter conclusion is reinforced by experiments on the oxidation of clean SiC(0001), which show that under the above oxidation conditions, the formation of $SiO_2$ proceeds without buildup of graphitic or amorphous carbon on the surface.

In addition to the changes in shape in the Si 2*p* and C 1*s* spectra upon oxidation, both spectra shift rigidly to lower binding energy. This is indicative of a band bending effect caused by the introduction of negative charge in the surface region. We compare this oxidation-induced band bending of the middle row of Fig. 2 with the band bending resulting from growing graphene multilayers on SiC made via SiC decomposition, shown in the top panel. We note that the band bending induced by adding electrically active graphene onto the surface, shown in the top panel, is similar to the band bending produced by low temperature oxidation. An obvious interpretation of this coincidence is that following oxidation the graphene has become electronically decoupled from the substrate, and has become electrically active, exhibiting roughly the same local chemical environment (e.g. doping level) as few-layer graphene. After oxidation, subsidiary experiments were performed which showed that flash heating to 1200 °C causes the oxide to decompose. Analysis of the C 1*s* spectra of these samples before and after flashing shows no significant difference in the intensities of the graphenic carbon peaks before and after flashing, indicating that the graphene layer is unaffected by this process, an observation we shall exploit below.

Selected-area low-energy electron diffraction (LEED) and *ex situ* AFM images, recorded from the same area of the surface before and after oxidation, are shown in Fig. 3.

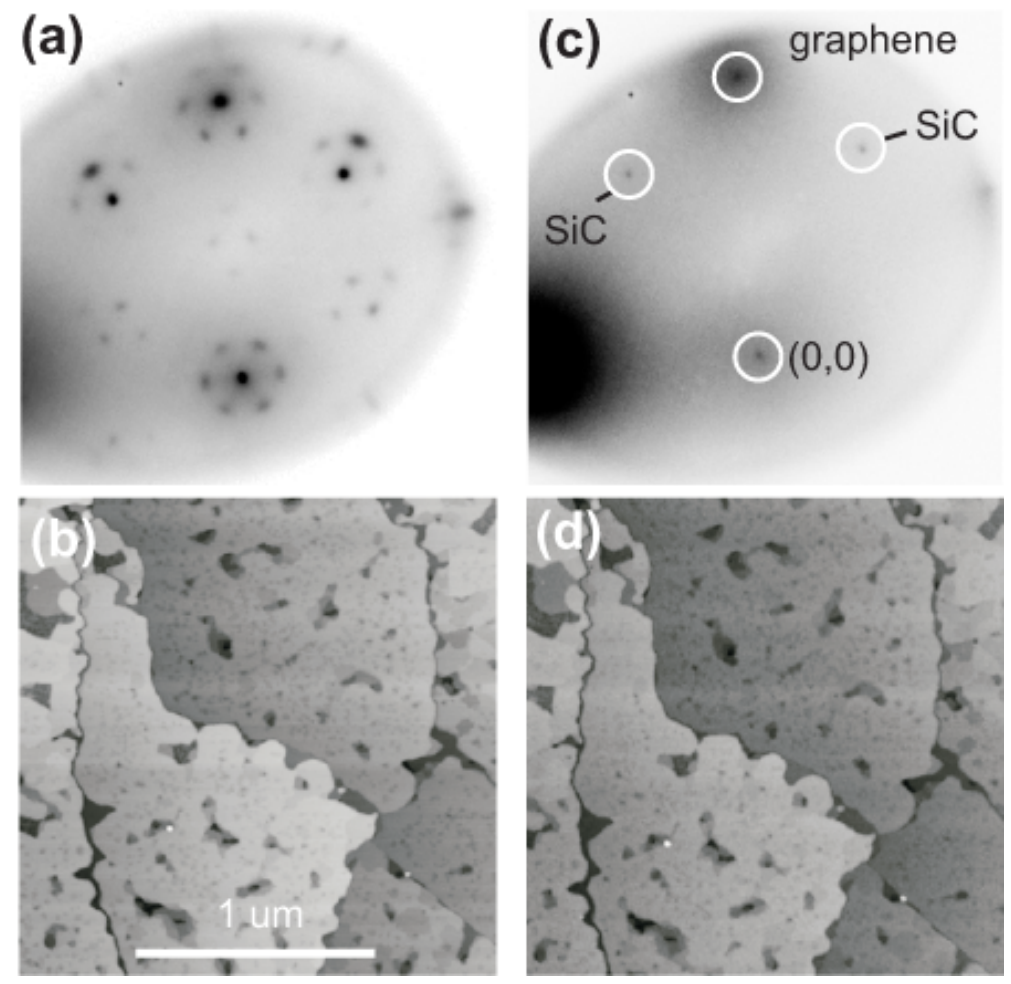

FIG. 3: LEED patterns (a,c) and AFM images (b,d) recorded from a graphene buffer layer on SiC(0001) before (a,b) and after (c,d) oxidation of the SiC substrate. The AFM images are nearly identical, while the diffraction pattern show that strong coupling to the SiC lattice is lifted by the oxidation.

The AFM images are virtually identical, demonstrating that the graphene is not consumed during the oxidation of the substrate. Note that the defect features, such as the small holes in the graphene layer (arising from the high nucleation rate of the graphene during CVD), have identical shapes and sizes before and after oxidation. This “edge graphene” would certainly be the most reactive feature of the buffer layer, and even it is apparently unaffected. In addition, we observe no features which could be attributed to silicon oxide on the surface (e.g. hillocks or protrusions), which suggests that the oxide formed is sub-surface, as desired. While AFM suggests no significant change in the surface morphology, the LEED patterns are quite different, suggesting a decoupling of the buffer layer from the substrate. Before oxidation, the expected $6\sqrt{3} \times 6\sqrt{3}$ diffraction pattern of the buffer layer is observed. The fractional-order spots arise from double diffraction from the SiC(0001) and graphene lattices[9], consistent with a strong coupling of the graphene buffer layer to the substrate. However, following oxidation, the fractional-order spots are extinguished. The pattern corresponds to a superposition of diffraction from graphene and from SiC(0001), indicative of a weaker coupling to the substrate, e.g. due to the formation of a thin amorphous silicon oxide layer between the graphene layer and the substrate. Similar changes in the LEED pattern are observed during H intercalation[17].

EELS spectra recorded after oxidation suggest that the buffer layer has adopted the electronic structure of graphene. The EELS spectrum (Fig. 1c)) exhibits a loss feature at 6.2 eV, where none was present before (Fig. 1b). This feature occurs at the same energy as the π-plasmon feature seen on the multilayer graphene flake (Fig. 1a). From this observation we conclude that the oxidation process has indeed decoupled the buffer layer from the substrate and restored the graphene-like π-bands.

The XPS and AFM data suggest that the oxidation process results in a very thin oxide layer under the buffer layer. In order to directly determine both the oxygen content and the location of the oxygen relative to the graphene, we performed structural measurements using medium energy ion scattering (MEIS)[24]. Using this highresolution form of Rutherford backscattering, we measured the depth profiles for oxygen, carbon, and silicon, verifying that the oxygen accumulates in a thin $SiO_2$ layer underneath the graphene. For these experiments SiC(0001) samples with a 1-2 graphene layers were prepared via sublimation and characterized using XPS. In Fig. 4 we show typical data for a sample before and after oxidation, taken using a normally incident beam of 100 keV protons. The data for each element have been replotted on a depth scale, and the intensities have been nornmalized to the cross sections.

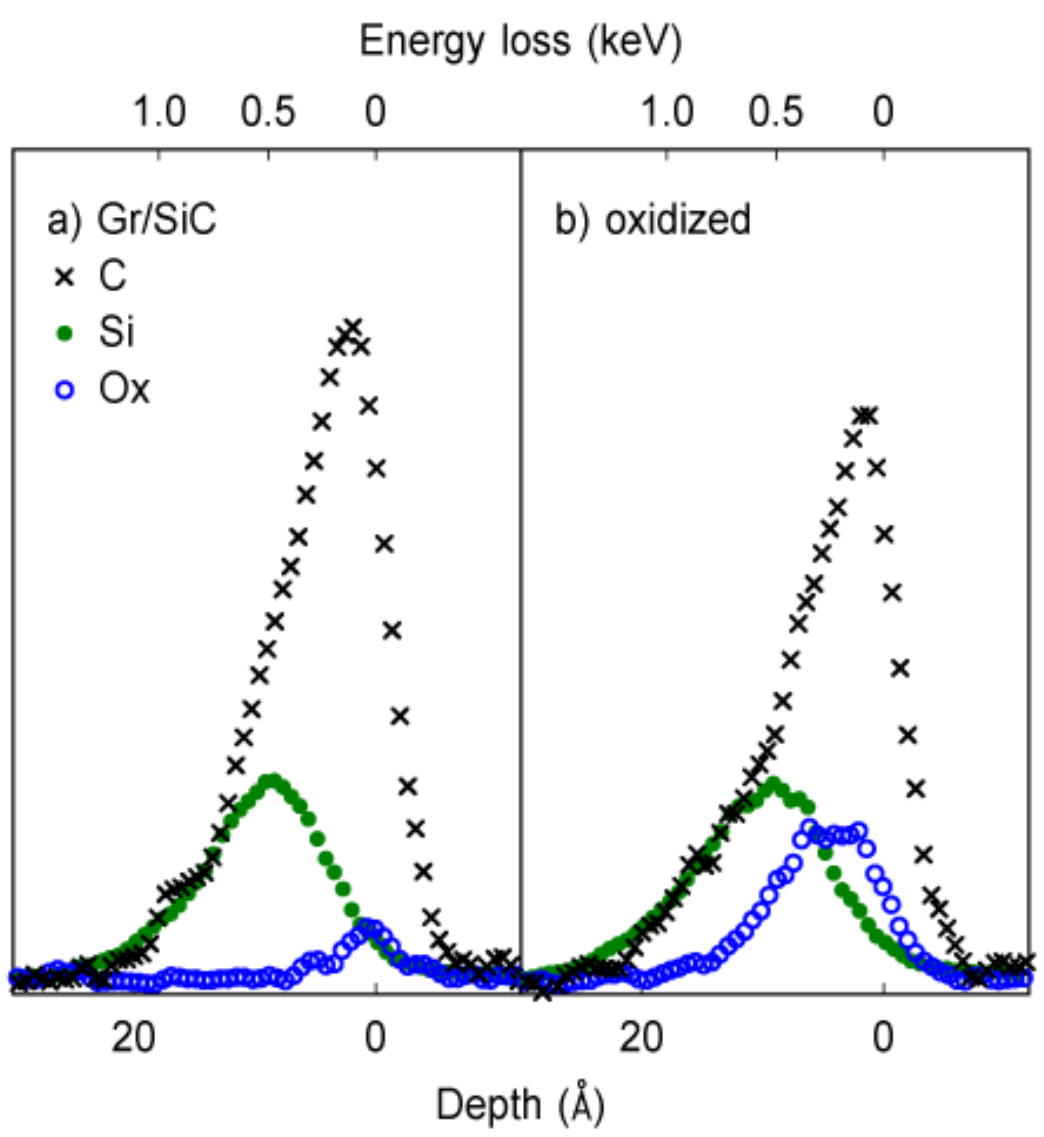

FIG. 4: Medium energy ion scattering spectra for graphene/SiC(0001) before (a) and after oxidation (b). Contributions from carbon, silicon and oxygen have been shifted and plotted on the same depth scale. After oxidation, the oxygen leading edge occurs deeper than the carbon edge, showing that the oxygen is subsurface.

Before oxidation (Fig. 4a), a large surface carbon peak is seen, caused by the graphene layer. For this sample, the graphene film thickness corresponded to roughly 2 ML (the buffer layer and one additional carbon layer). Deeper into the sample, the carbon intensity drops to the same level as the subsurface silicon peak, representing the contribution from the outermost layers of the SiC substrate. Only a small oxygen peak is observed, due to ambient exposure during transfer to the MEIS system. After oxidation (Fig. 4b), a more pronounced oxygen peak is observed. Note that the leading edge of the oxygen signal occurs deeper than the carbon edge, demonstrating that the oxygen located below the surface. The carbon peak has a slightly smaller intensity in the oxidized spectrum, which we attribute to non-uniformity in the graphene film thickness, consistent with XPS measurements which suggest that the thickness varies from 1-2 ML over this sample. Quantitative modeling of the spectrum supports the idea of an oxide-supported graphene film; we were able to accurately fit the results with of the unoxidzed sample as 1.9 layers of Gr/SiC(001) and after oxidation as 1.9 layers of Gr/ 3.4 Å $SiO_2$/ SiC(0001).

In summary, we have shown that the covalent bonding of the graphene buffer layer to the SiC(0001) substrate can be lifted by the insertion of an ultra-thin (~3 Å) oxide layer between the graphene and the substrate. The activated buffer layer exhibits the π-plasmon characteristic of graphene, showing that the band structure of graphene has been largely recovered. The lowtemperature oxidation method offers potential advantages for the device fabrication. It is simple to implement, can be carried out on pre-fabricated devices (i.e. with metal contacts in place), and is compatible with conventional CMOS processes. This work was supported by DARPA under contract FA8650-08-C-7838 through the CERA program.